# Upper Bound Estimate of the Electronic Scattering Potential of a Weakly-interacting Molecular Film on a Metal

Dhaneesh Kumar[†, §], Matthew Hendy[†], Jack Hellerstedt[†, §], Joanna Hewes[†], Shon Kolomoisky[†], Cornelius Krull[†, §], and Agustin Schiffrin[†, §, *].

[†]School of Physics & Astronomy, Monash University, Clayton, Victoria 3800, Australia

[§]ARC Centre of Excellence in Future Low-Energy Electronics Technologies, Monash University, Clayton, Victoria 3800, Australia

* agustin.schiffrin@monash.edu

ABSTRACT: Thin organic films and two-dimensional (2D) molecular assemblies on solid surfaces yield potential for applications in molecular electronics, optoelectronics, catalysis and sensing. These applications rely on the intrinsic electronic properties of the hybrid organic/inorganic interface. Here, we investigate the energy dispersion of 2D electronic states at the interface between an atomically thin self-assembled molecular film – comprised of flat, non-covalently bonded 9,10-dicyanoanthracene (DCA) molecules – and a Ag(111) surface. Using Fourier-transformed scanning tunnelling spectroscopy (FT-STS), we determined that the 2D electronic wavefunctions with wavevectors within ~80% of the first Brillouin zone (BZ) area close to the Γ-point are free-electron-like, suggesting a weak electronic interaction between the 2D molecular film and metal surface. Via a perturbative 2$^{nd}$ order correction to the free electron energy dispersion, we further established an upper bound for the amplitude of the scattering potential resulting from the self-assembled molecular film, that the interface electrons are subject to, on the order of 1.5 eV. Our approach allows for quantifying electronic interactions at hybrid 2D interfaces and heterostructures.

INTRODUCTION: Thin organic molecular films on metals are promising for electronic, optoelectronic, catalytic and sensing applications.[1–7] The realization of such applications relies on fundamental understanding and characterisation of electronic properties and interactions at the metal-molecule interface.[8–10] While electronic hybridisation, metal-to-molecule charge transfer, and adsorption-induced changes of molecular conformation can dramatically alter the intrinsic electronic properties of the organic films,[11] the latter can also affect the inherent chemistry and electronics of the metal surface. In particular, electronic states located at the surface of metals, and delocalized in two-dimensions (2D) along this surface (i.e., with dispersive eigenenergies) – such as Shockley surfaces states or hybrid interface states[12–15] – can be scattered and even confined (e.g., to 1D or 0D; along or perpendicular to the surface) by adsorbed molecular nanostructures or interfacing organic films, potentially leading to, e.g., electron eigenenergy shifts, distortion of electron eigenenergy dispersions (i.e., changes in electron group velocities), and/or electronic localization.[13,16,17] These 2D surface/interface states can have eigenenergies lying near or at the metal Fermi level [e.g., 2D free-electron-like Shockley states at (111) surfaces of noble metals], and can significantly influence charge carrier dynamics across the interface.[18,19] For example, such interface states, owing to their large wavefunction overlap with metallic and organic electronic states, can lead to more efficient electron transfer across the interface.[20]

Electronic states at hybrid organic-metal interfaces can be investigated by photoelectron spectroscopy (PES) techniques, e.g., ultraviolet (UPS)[21,22], x-ray (XPS), angle-resolved[23–25] (ARPES) or two-photon PES[14,19,26,27], allowing for determining density of states and 2D energy dispersions. However, these techniques offer spatially averaged information (hindering the observation of local phenomena given by e.g., defects, different nano-/meso-scopic domains) with limited energy resolution (at best, ~5 meV, typically above), and with UPS, XPS and ARPES only probing occupied states.

Alternatively, scanning tunnelling microscopy (STM) and spectroscopy (STS) allow for measuring the local density of electronic states (LDOS), with atomic spatial resolution and sub-meV energy resolution. In particular, Fourier-transformed scanning tunnelling spectroscopy (FT-STS)[28–32] provides information on the energy dispersion, $E(\mathbf{k})$, of electronic states (both occupied and unoccupied) at surfaces and interfaces, as a function of wavevector $\mathbf{k}$, while still being site-specific at the nanoscale thanks to the real-space resolution of STM and STS. For example, this technique has been shown to elucidate the electronic structure of superconductors,[30–32] many-body effects on the Shockley surface state of Ag(111)[15], topological surface states[33], etc. However, FT-STS studies of the electronic properties of 2D hybrid organic/metallic interfaces remain limited[12,13,34,35]. In particular, topographic and electronic contributions from 2D organic overlayers are inevitably intertwined in FT-STS measurements, challenging the unambiguous characterization of electronic properties of hybrid organic/metallic interfaces via this method[12].

Such measurements are often limited to (quasi-)elastic scattering of 2D electronic states, where energy dispersions of such states are easy to access and are mainly parabolic in nature[13,34,35].

Here, we use FT-STS to study the 2D electronic states at the hybrid organic/metal interface[36] composed of the 9,10-dicyanoanthracene (DCA) molecular self-assembly on Ag(111). We find that this interface state (IS) consists of the underlying Ag(111) Shockley surface state upshifted in energy via vertical confinement by the adsorbed DCA self-assembly. This 2D IS is characterized by a parabolic energy dispersion for crystal momenta **k** close to the Γ-point within ~80% of the first Brillouin zone (BZ) area of the 2D DCA monolayer, indicative of a free-electron-like behavior and of a weak interaction between the 2D molecular self-assembly and noble metal surface. Using $2^{nd}$ order corrections for the energy dispersion of electrons in a weak 2D periodic potential – imposed by the molecular film –, we find slight deviations from parabolic behavior for wavevectors close to the BZ boundary. From this, we estimate an upper bound for the amplitude of the electronic scattering potential resulting from the 2D molecular self-assembly, on the order of 1.5 eV. Although scattering of Shockley surface states of noble metal (111) surfaces by flat aromatic molecules has been studied by STS previously[37], the upper bound of the electronic scattering potential magnitude has not been discussed explicitly so far.

METHODS

*Sample preparation.* We obtained DCA dimers (Figure 1a) by depositing DCA molecules (Tokyo Chemical Industry; >95% purity) from the gas phase onto a clean Ag(111) surface held at 4.4 K, in ultrahigh vacuum (UHV). Self-assembled DCA sub-monolayers (Figures 2-5) were synthesised similarly, but with the clean Ag(111) surface held at room temperature. The base pressure during molecular deposition was below $5 \times 10^{-10}$ mbar. DCA was sublimed at 373 K resulting in a deposition rate of ~0.001 nm/s. The surface density of DCA dimers is proportional to deposition time. The size of the DCA sub-monolayer domains are limited by the area of the bare Ag(111) terraces, and by the molecular coverage and deposition time. The Ag(111) surface was cleaned by repeated cycles of $Ar^+$ sputtering and annealing at 790 K.

*STM & STS measurements.* All STM and STS measurements were performed at 4.4 K in UHV ($\sim 1 \times 10^{-10}$ mbar) with an Ag-terminated Pt/Ir tip. All topographic STM images were taken in constant-current mode with the sample bias reported throughout the text. All d$I$/d$V$ STS spectra, unless otherwise stated, were acquired by recording the tunnelling current as a function of tip–sample bias voltage in the junction. During these STS measurements, the tip–sample distance was stabilized with respect to a specified tunnelling current setpoint. We then numerically differentiated the resulting $I$–$V$ data to obtain d$I$/d$V$ as a function of tip–sample bias voltage. All d$I$/d$V$ maps required for FT-STS (Figures 3, 4) were acquired in constant-current mode (in order to be more sensitive to real-space LDOS changes relative to the topographic corrugation; $I_t = 2$ nA) with a lock-in technique, by applying a 1.13 kHz modulation to the bias voltage. The amplitude of this lock-in bias modulation, $V_{mod}$, depended on the d$I$/d$V$ map acquisition bias (e.g., $V_{mod} = 1$ and 15 mV for $V_b = 230$ and 600 mV, respectively), compensating for the weakening scattering signal at higher energies. d$I$/d$V$ maps with $V_b \leq 350$ mV were obtained over a $124 \times 124$ nm$^2$ defect-filled DCA/Ag(111) region with a spatial resolution of $\sim 0.37$ nm/pixel (e.g., Figure 3b). Maps with $V_b > 350$ mV were obtained over an $83 \times 83$ nm$^2$ region with a spatial resolution of $\sim 0.35$ nm/pixel (e.g., Figure 3d). This ensured a **q**-space resolution of $\Delta \mathbf{q} \sim 0.05$ nm$^{-1}$ and $\Delta \mathbf{q} \sim 0.08$ nm$^{-1}$ for $V_b \leq 350$ mV (where scattering is dominant at low $|\mathbf{q}|$) and for $V_b > 350$ mV (where the scattering is dominant at high $|\mathbf{q}|$), respectively. This further ensured a reliable extraction of scattering wavevectors, with $|\mathbf{q}| < 8.5$ nm$^{-1}$ and $|\mathbf{q}| < 9.0$ nm$^{-1}$ for maps with $V_b \leq 350$ mV and $V_b > 350$ mV, respectively (according to the Nyquist-Shannon sampling theorem[38]). Real-space d$I$/d$V$ maps in Figures 3c, d were bandpass-filtered, keeping Fourier components $1.4 < |\mathbf{q}| < 2.3$ nm$^{-1}$ and $3.1 < |\mathbf{q}| < 4.5$ nm$^{-1}$, respectively. In order to avoid possible anisotropic distortions of **q**-space FT-STS d$I$/d$V$ maps in Figure 4, and to accurately determine the energy dispersion $E(\mathbf{k})$, we scale calibrated (in the x-y plane) the real-space d$I$/d$V$ maps using the known registration of the DCA molecule film with respect to the underlying Ag(111) substrate.[36] The **q**-space d$I$/d$V$ line profiles (e.g., Figure 4b) enabling the extraction of $E(\mathbf{k})$ were radially averaged within an angle window increased with increasing acquisition bias voltage, to increase signal-to-noise ratio (e.g., ±1° at $V_b$ = 120 mV; ±5° at $V_b$ = 1.5 V). Defects (consisting of displaced molecules and required for FT-STS; see Figure 3a) were introduced into large, pristine self-assembled DCA domains (> 130 × 130 nm$^2$) by scanning with parameters $V_b$ = -3 V, $I_t$ = 3.5 nA (i.e., resulting in significant tip–sample interaction). This process is reversible, i.e., the displaced molecules can regain their initial position within the film *via* lateral STM manipulation.

*Goodness of fits.* Throughout the text, we calculated chi-squared, $\chi^2$, as a measure of the goodness of fit to our data:

$$\chi^2 = \frac{1}{N} \sum_i \frac{(y_i^{(\text{model})} - y_i^{(\text{data})})^2}{\delta y_i^2}$$

where $y_i^{(\text{data})}$ and $\delta y_i^2$ refer to the $i^{\text{th}}$ data value and the uncertainty associated with that data value, respectively; $y_i^{(\text{model})}$ refers to the $i^{\text{th}}$ fit model value; $N$ is the total number of data points.

*Error bars.* All error bars reported throughout the text correspond to a 95% confidence interval except for the energy dispersion data, $E(\mathbf{k})$ (e.g., Figure 4d). The uncertainty in **k** for the dispersion data $E(\mathbf{k})$ was estimated by combining in quadrature the half width at half maximum (HWHM) of the Lorentzian functions used to fit the scattering Fourier peaks (e.g., in Figure 4b) with the uncertainty in energy given by the lock-in bias modulation amplitude, $V_{mod}$:

$$\delta k^2 = (\text{HWHM})^2 + \left(\frac{m^*\delta E}{\hbar^2 k}\right)^2$$

with[39] $\delta E = V_{mod}/\sqrt{6}$. The second term above was derived from $\delta E/(dE/dk)$ with:

$$\frac{dE}{dk} = \frac{d}{dk}\left(\frac{\hbar^2 k^2}{2m^*}\right) = \frac{\hbar^2 k}{m^*}$$

RESULTS AND DISCUSSION

Figure 1a shows an STM image of a DCA dimer (i.e., two DCA molecules linked *via* non-covalent proton acceptor-ring interactions[40] between cyano nitrogen atoms and hydrogen atoms of adjacent molecule) after deposition of DCA on a Ag(111) surface held at 4.4 K (see Methods for sample preparation details). Each bright elliptical feature corresponds to a single DCA molecule, adsorbed in a planar configuration[36] (see Figure 1a inset). At this low temperature, the DCA molecules are able to diffuse on the surface, enabling the non-covalent intermolecular interaction and formation of such dimers. This indicates a small adsorption energy and hence weak interaction between DCA and noble metal, consistent with previous work.[36] In Figure 1a (acquired at a small bias voltage $V_b = -20$ mV) we further observe a circular interference pattern around the DCA dimer, attributed to Friedel oscillations[12,41,42] in the Ag(111) LDOS, resulting from scattered Shockley surface state electron wavefunctions off the DCA dimer.[15,43]

Figure 1b displays the STM apparent height profile, $z_{app}$, across the DCA dimer in Figure 1a. For distances $|x| < 2.5$ nm (cyan region) from the dimer center, $z_{app}$ reflects the topography of

the DCA dimer. For $|x| > 2.5$ nm, $z_{app}$ shows the symmetric (about $x = 0$ nm), decaying Friedel oscillation of the near-Fermi LDOS, which for point-like defects can be described by:[44]

$$\Delta \text{LDOS}(k_0, x) = \frac{A\left[\cos^2\left(k_0|x| - \frac{\pi}{4} + \partial_0\right) - \cos^2\left(k_0|x| - \frac{\pi}{4}\right)\right]}{k_0|x|} \quad (1),$$

where $A$ is the amplitude, $\partial_0$ is the scattering phase shift, and $k_0 = \sqrt{2m^*(E - E_0)}/\hbar$, with $m^*$ and $E$ being the surface state electron's effective mass and energy, respectively, and $E_0 \approx -65$ mV being the onset energy of the Ag(111) surface state energy dispersion.[15,45] The fit of $z_{app}(x)$ in Figure 1b with Eq. (1) yields $m^*/m_e = 0.421 \pm 0.001$, consistent with previous works[15,45], and $\partial_0 = -51.8 \pm 4.0°$.

In the following, we focus on estimating the strength of this scattering potential via differential conductance ($dI/dV \propto$ LDOS) STS, in the particular case where Ag(111) is covered by self-assembled domains of non-covalently bonded DCA molecules. Figure 2a shows an STM image of one such domain, with molecular self-assembly unit cell vectors $\mathbf{a_1}$ and $\mathbf{a_2}$, after deposition of DCA on Ag(111) held at room temperature (see Methods). Further information on the self-assembly of DCA on Ag(111) can be found elsewhere.[36]

Figure 2b shows $dI/dV$ STS spectra measured across the boundary of a self-assembled DCA domain. As previously reported,[36] spectra at the DCA anthracene extremity show a peak-like feature at $V_b \approx 350$ mV corresponding to the DCA lowest unoccupied molecular orbital (LUMO). We observe a continuous evolution of the bare Ag(111) $dI/dV$ spectrum—with a step-like feature with energy onset ~-65 mV, associated with the LDOS of the Shockley surface state 2D free electron gas (2DEG), and with oscillations resulting from scattering of the latter by the molecular domain boundary—to the $dI/dV$ spectrum taken at the DCA centre within the molecular domain showing a step-like feature with energy onset ~120 mV. From this, we associate the ~120 mV onset with a 2DEG at the DCA-Ag(111) interface, resulting from vertical confinement of the Ag(111) Shockley surface state by the molecular film (i.e., upshifting the onset energy by ~200 mV).

Similar 2D electronic interface states (IS) have been observed for a variety of molecule-metal systems,[12,13,35,46,47] with different energy onsets. The evolution of the noble metal Shockley surface state to such an IS can be modelled via a single parameter, the adsorption height, $d_C$, of the organic overlayer,[48] within the assumption that these molecular overlayers are physiosorbed (i.e., weak molecule-metal chemical interaction). With[36] $d_C$ = 2.85 ± 0.05 Å (inferred via STM) for DCA on Ag(111), this simple model predicts an IS energy onset ~190 mV larger than that of the Ag(111) Shockley surface state, in excellent agreement with our measured IS energy onset shift (~200 mV). This is consistent with our previous findings that DCA interacts weakly with Ag(111), and confirms that the IS results from the vertically confined, energy-upshifted Ag(111) Shockley surface state.

To gain further insight into this IS, we performed FT-STS measurements[15,28,49] by introducing defects to the DCA molecular film that act as scatterers of the IS electrons. These scatterers consist of molecules displaced controllably from their initial position by the STM tip (Figure 3a; see Methods for details). In the d$I$/d$V$ maps ($V_b$ = 230 mV) in Figures 3b, c, we observe around each defect an isotropic modulation of the d$I$/d$V$ signal. This d$I$/d$V$ signal modulation is observed for bias voltages $V_b$ above the IS energy onset of ~120 mV (Figure 2); we attribute it to Friedel oscillations of the LDOS caused by interfering incident and scattered IS wavefunctions. Similarly, Figure 3d displays the bandpass-filtered (see Methods) d$I$/d$V$ map for $V_b$ = 600 mV. The period of the LDOS Friedel oscillations decreases with increasing bias voltage.

Fourier transforms (FT) of these d$I$/d$V$ maps allow us to extract the scattering wavevector, **q**, associated with electronic wavefunctions with a specific eigenenergy (i.e., given by the d$I$/d$V$ map acquisition bias voltage $V_b$ with respect to the metal surface Fermi level).[15,28,49] Figure 4a shows the FT of the d$I$/d$V$ map in Figure 3b. We observe FT peaks corresponding to the molecular film periodicity (cyan dashed circles). Additionally, we observe a prominent isotropic ring-like feature (white arrow) about |**q**| = 0 nm$^{-1}$ whose radius is related to the real-space period of the LDOS modulation due to scattering of the IS by defects in Figures 3b, c. [15,28,49]

We determined the **q**-space locations of the IS scattering features via Lorentzian fitting of the FT intensity profile along high-symmetry directions (e.g., X′-Γ-Y′; Figure 4b), at a given bias voltage.

For increasing bias voltage, the Fourier space radius of the scattering ring-like feature increases (Figure 4c), i.e., the real-space period of the scattering-induced Friedel oscillations decreases (Figure 3). By extracting the **k**-space radius of these IS-scattering-related ring features (i.e., Lorentzian fitting as in Figure 4b) for different bias voltages, we determined the energy dispersion $E(\mathbf{k})$ (e.g., along X-Γ-Y in Figure 4d; see Supporting Information [SI] for data along all high-symmetry directions). Here, the electronic wavefunction wavevector **k** and the scattering wavevector **q** are related by the relation **q** = 2**k**; that is, an electron characterized by a wavefunction with initial wavevector **k** can be scattered (e.g., by an impurity or defect) into a final state associated with a wavevector **k**′, following **q** = **k**′ − **k**, where **q** is the scattering wavevector. Since electron scattering is most likely dominated by elastic back-scattering events, we have that **k**′ = −**k**, hence **q** = 2**k**. We fit $E(\mathbf{k})$ along all the high-symmetry directions (i.e., along Γ-X, Γ-H1, Γ-C, Γ-H2, and Γ-Y) with a free electron dispersion relation given by:

$$E^{(0)}(\mathbf{k}) = \hbar^2|\mathbf{k}|^2/2m_\mathbf{k}^* + \mu \qquad (2),$$

Here $m_\mathbf{k}^*$ is the IS electron effective mass along the $\mathbf{k}$ direction and $\mu$ is the IS energy onset. This fit (e.g., black curve in Figure 4d) yields $\mu = 122 \pm 1$ mV, and $m_\mathbf{k}^*/m_e = m^*/m_e = 0.395 \pm 0.005$ independent on the direction of $\mathbf{k}$ ($m_e$: electron mass), similar to the effective mass for the bare Ag(111) Shockley surface state,[15] with a fit chi-squared $\chi^2_{Eq(2)} = 0.098$ (see Methods).

The energy dispersion $E(\mathbf{k})$ can deviate from the free electron-like behaviour (i.e., parabolic dispersion) when $\mathbf{k}$ approaches the BZ boundaries of the molecular film (e.g., X and Y in Figure 4d; note subsequent mentions of BZ refers to that of molecular film unless otherwise stated), where IS electrons experience most the scattering potential imposed by the 2D molecular self-assembly. In the following, we address whether perturbative corrections to the free electron dispersion provide a better fit (i.e., smaller $\chi^2$) of our experimental $E(\mathbf{k})$ when $\mathbf{k}$ approaches the BZ boundary. We consider (along all high-symmetry directions) a nearly-free electron dispersion relation including a 2nd order correction to $E^{(0)}(\mathbf{k})$ in Eq. (2):[50]

$$E(\mathbf{k}) = E^{(0)}(\mathbf{k}) + \sum_\mathbf{G} \frac{|U_\mathbf{G}|^2}{(E^{(0)}(\mathbf{k}) - E^{(0)}(\mathbf{k}-\mathbf{G}))} \quad (3),$$

where $U_\mathbf{G}$ are the Fourier coefficients of the electronic scattering potential resulting from the periodic DCA molecular film:

$$U(\mathbf{r}) = \sum_\mathbf{G} U_\mathbf{G} e^{i\mathbf{G}\cdot\mathbf{r}} \quad (4),$$

We assume that in Eq. (4), the only non-zero Fourier coefficients, $U_\mathbf{G}$, are those associated with $\mathbf{G} = \{\pm\mathbf{G_1}, \pm\mathbf{G_2}, \pm(\mathbf{G_1}+\mathbf{G_2})\}$, where $\mathbf{G_1} = G_{1x}\hat{\mathbf{x}} + G_{1y}\hat{\mathbf{y}}$ and $\mathbf{G_2} = G_{2x}\hat{\mathbf{x}} + G_{2y}\hat{\mathbf{y}}$ are the primitive unit cell vectors of the DCA molecular film reciprocal lattice, related to real-space lattice vectors $\mathbf{a_1} = a_{1x}\hat{\mathbf{x}} + a_{1y}\hat{\mathbf{y}}$ and $\mathbf{a_2} = a_{2x}\hat{\mathbf{x}} + a_{2y}\hat{\mathbf{y}}$ in Figure 3a as $\begin{pmatrix} G_{1x} & G_{2x} \\ G_{1y} & G_{2y} \end{pmatrix} =$

$\frac{2\pi}{a_{1x}a_{2y}-a_{1y}a_{2x}}\begin{pmatrix} a_{2y} & -a_{1y} \\ -a_{2x} & a_{1x} \end{pmatrix}$. These Fourier coefficients have the largest contribution to the 2D molecular potential. With $U_{\mathbf{G}} = U_{-\mathbf{G}}$ due to symmetry, $U(\mathbf{r})$ becomes:

$$U(\mathbf{r}) \approx 2U_{\mathbf{G}_1}\cos(\mathbf{G}_1 \cdot \mathbf{r}) + 2U_{\mathbf{G}_2}\cos(\mathbf{G}_2 \cdot \mathbf{r}) + 2U_{\mathbf{G}_1+\mathbf{G}_2}\cos((\mathbf{G}_1 + \mathbf{G}_2) \cdot \mathbf{r}) \quad (5),$$

We fit $E(\mathbf{k})$ with Eq. (3) for $\mathbf{k}$ along all the high-symmetry directions (i.e., along Γ-X, Γ-H1, Γ-C, Γ-H2, and Γ-Y; see SI Fig. S1), with parameters $m_{\mathbf{k}}^*$, $\mu$, $U_{\mathbf{G}_1}$, $U_{\mathbf{G}_2}$, and $U_{\mathbf{G}_1+\mathbf{G}_2}$ as fitting parameters. Based on the fitting of $E(\mathbf{k})$ with Eq. (2), $m_{\mathbf{k}}^* = m^*$ was assumed isotropic. The resulting fit (e.g., blue curve in Figure 4d) yields $\chi^2_{\text{Eq}(3)} = 0.074$, improving the fit given by a free electron dispersion. To determine to what extent the IS electrons behave like free electrons, we fit our experimental data, $E(\mathbf{k})$, with both free- and nearly-free-electron models, along all high-symmetry directions (i.e., along Γ-X, Γ-H1, Γ-C, Γ-H2, and Γ-Y), whilst considering wavevectors $\mathbf{k}$ whose magnitude $|\mathbf{k}|$ are below a specified threshold, $k_{\max}$. We then calculated $\chi^2$ as a function of $k_{\max}$ for both model fits (see Figure 4e). These $k_{\max}$ values are related to a percentage of the Brillouin zone area (for the DCA overlayer and centered about Γ), $\sim \frac{\pi k_{\max}^2}{A_{\text{BZ}}}$, where $A_{\text{BZ}}$ is the first Brillouin zone area. By comparing the fits of both models [i.e., Eq. (2) vs Eq. (3)], we find that the IS electronic dispersion is free-electron-like for $\mathbf{k}$ within ~80% of the first BZ area (centred about the Γ point). However, the experimental dispersion data agrees better with Eq. (3) than with Eq. (2) for $|\mathbf{k}| > 3$ nm$^{-1}$ (i.e., closer to the BZ boundary; left plot in Figure 4d). Indeed, for fits of our whole dispersion data (i.e., $|\mathbf{k}| > 0$ nm$^{-1}$) using Eq. (3), but with $\chi^2$ calculated only considering wavevectors with $|\mathbf{k}| > 3$ nm$^{-1}$ (again along all high-symmetry directions, i.e., along Γ-X, Γ-H1, Γ-C, Γ-H2, and Γ-Y), we obtained $\chi^2_{\text{Eq}(3)} = 0.05$, significantly better than $\chi^2_{\text{Eq}(2)} = 0.082$. This is further evidenced in Figure 4e, where $\chi^2_{\text{Eq}(3)} < \chi^2_{\text{Eq}(2)}$ for $k_{\max} > \sim 3.2$ nm$^{-1}$. Importantly, we note that the bare

Ag(111) Shockley surface state dispersion is known to deviate from the free-electron-like quadratic expression,[51] but at much higher energies and for $|\mathbf{k}| > 5$ nm$^{-1}$. Here, we are interested in FT-STS data for $0 < |\mathbf{k}| < 4$ nm$^{-1}$, where the energy dispersion of the bare Ag(111) Shockley surface state is free-electron-like and is well approximated by a quadratic function of $|\mathbf{k}|$.

The fit of $E(\mathbf{k})$ using Eq. (3) yielded $m^*/m_e = 0.385 \pm 0.005$, $|U_{\mathbf{G}_1}| = 0.29 \pm 0.05$ eV, $|U_{\mathbf{G}_2}| = 0.18 \pm 0.04$ eV, and $|U_{\mathbf{G}_1+\mathbf{G}_2}| = 0.17 \pm 0.05$ eV. These values represent an upper bound to the magnitude of the Fourier coefficients of $U(\mathbf{r})$. Dispersion data at higher energies ($V_b \geq 1.4$ V) can further constrain these values; however, extraction of such data is challenging given the severe attenuation of the d$I$/d$V$ scattering intensity at energies far above the IS onset (see SI Figure S2).

The obtained Fourier coefficients allow us to reconstruct an upper bound to the 2D molecular film potential $U(\mathbf{r})$ according to Eq. (5); see Figure 5a. Note that, although it is not trivial to unambiguously determine the sign of these Fourier components (i.e., related to whether the scattering potential is attractive or repulsive),[52] we can, however, estimate an upper bound of the magnitude of $U(\mathbf{r})$; we therefore set this sign as positive (i.e., repulsive scattering potential), as suggested by the scattering pattern in Figure 1 and its fit with Eq. (1). The reproduced molecular potential in Figure 5a is similar to STM imaging of the DCA molecular film (Figure 3a). This map of $U(\mathbf{r})$ shows variations of the electronic energy landscape, $\Delta U(\mathbf{r}) = \max[U(\mathbf{r})] - \min[U(\mathbf{r})]$, given by the molecular film that could reach $\Delta U(\mathbf{r}) \approx 1.5$ eV (Figure 5a inset).

We further calculated the 2D band structure $E(\mathbf{k})$ of the DCA/Ag(111) IS electrons using the nearly-free electron approximation.[50] We solved the central equation for $E(\mathbf{k})$:

$$\left(\frac{\hbar^2|\mathbf{k}|^2}{2m^*} - E(\mathbf{k})\right) C_\mathbf{k} = \sum_\mathbf{G} U_\mathbf{G} C_{\mathbf{k}-\mathbf{G}} \quad (6),$$

with $C_\mathbf{k}$ being the Fourier coefficients of the IS electronic wavefunctions (Figure 5b; see SI). Although the amplitude of $U(\mathbf{r})$ can be on the order of ~1.5 eV, in 2D, the bandgaps of $E(\mathbf{k})$ at the Brillouin zone boundaries are given by $U_\mathbf{G}$ along a particular $\mathbf{G}$ direction; here these gaps are at most on the order of ~0.5 eV. The shape and strength of $U(\mathbf{r})$ is crucial to the accurate determination of the band structure. Where piecewise constant potentials can be a good first approximation for a given system [e.g., in reproducing $E(\mathbf{k})$],[53] they may not necessarily reflect the actual potential energy landscape of the system. A more accurate derivation of $U(\mathbf{r})$ for our system would require a larger collection of Fourier components $\{U_\mathbf{G}\}$ than the one we considered. This could arguably be achieved with more FT-STS data at higher energies.

Alternative to our approach here,[12] $U(\mathbf{r})$ could also be reconstructed via subtle d$I$/d$V$ signatures of Bragg scattering of IS electrons (i.e., scattering of IS electrons characterized by wavevectors close to the BZ boundary, concomitant with modulations of the d$I$/d$V$ spectra, with subtle d$I$/d$V$ maxima at energies associated with band edges due to gap openings at the BZ boundary). We note that within the energy window considered here (from 0.2 to 1.4 eV, with the d$I$/d$V$ signal related to the IS scattering significantly reduced for larger energies; see SI Figure S2), the IS energy dispersion does not cross any BZ boundary (see Figure 5b), and hence d$I$/d$V$ signatures of Bragg scattering by the molecular film were not observed. Our approach, however, offers an alternative for estimating the maximum possible magnitude of the electronic scattering potential resulting from atomic or molecular adsorbates in the case where the latter interact weakly with the surface and where d$I$/d$V$ signatures of Bragg scattering are not observed.

Our method provides a direct and quantitative upper bound estimate of the plausible deviation of the IS energy dispersion from free electron (parabolic) behaviour as $\mathbf{k}$ approaches the BZ boundary. In particular, our approach is applicable to similar physiosorbed 2D molecular films where the formation of an IS can be ascribed mostly to a 2D surface state of the underlying surface upshifted in energy due to electronic confinement. By evaluating the goodness of the 2$^{nd}$-order nearly-free-electron-model fit (i.e., $\chi^2$ value) of the FT-STS energy dispersion data, for different **k**-ranges away from the Γ-point along high-symmetry directions of the 2D molecular film, we can address how likely IS electrons are to behave differently from free particles. In our work, the $\chi^2$ values assuming either the free- or nearly-free-electron-model are comparable up to ~80% of the BZ range close to Γ. Beyond this range, $\chi^2$ for the 2$^{nd}$-order nearly-free-electron-model fit is significantly smaller (i.e., better fit) than the free electron model (Figure 4e). Notably, our approach provides an upper bound estimate of $U(\mathbf{r})$ without the requirement to obtain FT-STS data for **q**-vectors associated exactly with the BZ boundary, where **q**'s coincide with structural Fourier peaks given by the 2D molecular film periodicity (given that $\mathbf{q} = 2\mathbf{k}$; Figure 4a). The latter results in a distortion of the **k**-space electronic scattering pattern and challenges an accurate evaluation of $E(\mathbf{k})$ at energies associated with **k**'s belonging to the BZ boundary.

It is worth noting that for the DCA self-assembly on Ag(111) here, although the electron energy dispersion $E(\mathbf{k})$ is mostly quadratic for a vast majority of **k**'s in the BZ (i.e., ~80% of the BZ range close to Γ along high-symmetry directions), the amplitude of $U(\mathbf{r})$ – to which surface electrons are exposed and that results from the 2D molecular film – can still be very significant (~1.5 eV), even when the interaction between molecular film and underlying surface appears relatively weak.[36]

CONCLUSIONS

To conclude, we have shown the presence of a delocalized 2DEG at the interface between a 2D self-assembly of DCA molecules and a Ag(111) surface, resulting from vertical confinement of the Ag(111) Shockley surface state by the molecular film. By FT-STS, we derived the IS energy dispersion, $E(\mathbf{k})$. Using $2^{nd}$ order corrections to the free-electron dispersion, we inferred an upper bound for the real-space 2D electrostatic potential, $U(\mathbf{r})$, imposed by the DCA molecular film on the IS electrons. This upper bound estimate is associated with a potential variation, $\Delta U(\mathbf{r})$, of ~1.5 eV. Our work provides a viable tool for probing 2D electrostatic potentials and electronic energy landscapes at hybrid organic-inorganic interfaces, even between weakly interacting materials.

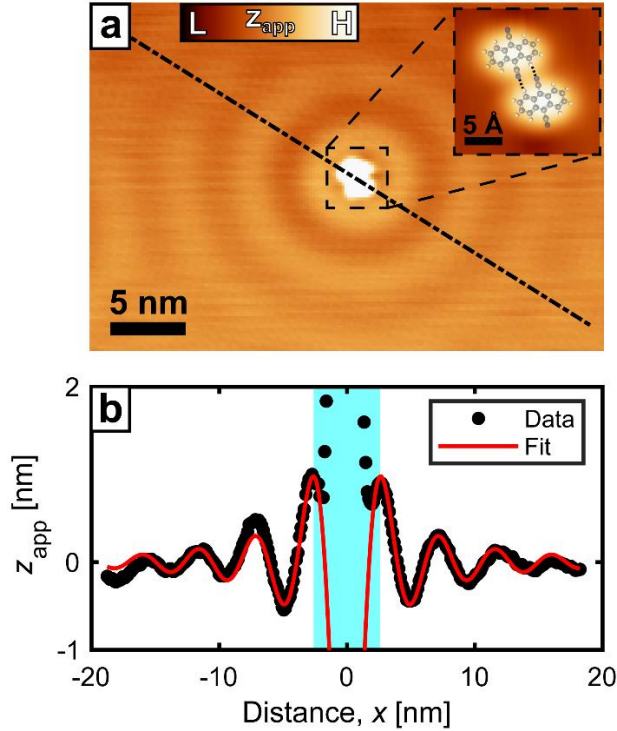

**Figure 1.** (a) Constant-current STM image of DCA dimer on Ag(111)($V_b$ = -20 mV, $I_t$ = 50 pA). Each bright elliptical feature represents one DCA molecule, linked to the other molecule via non-covalent interactions between nitrogen and hydrogen (see inset with overlaid ball-and-stick model of DCA; grey: carbon; blue: nitrogen; white: hydrogen). Scattering of Ag(111) Shockley surface state off DCA dimer results in circular interference pattern (i.e. Friedel oscillations) around DCA dimer. (b) STM apparent height profile (black dots) across DCA dimer in (a) (black dashed line). Red curve: fit of apparent height profile given by Eq. (1). Cyan region corresponds to topography of DCA dimer, which was not considered in the fit.

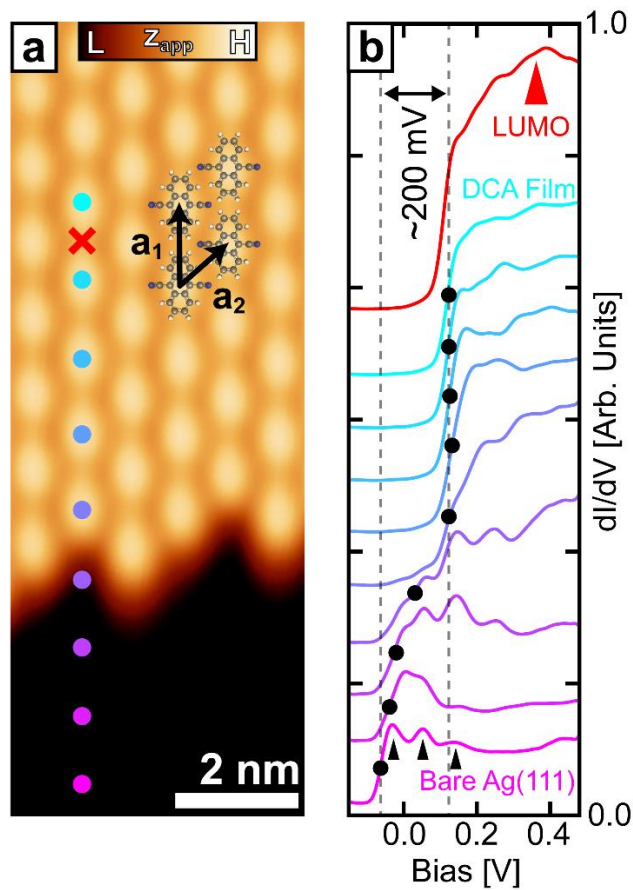

**Figure 2.** (a) Constant-current STM image of a self-assembled DCA molecular domain on Ag(111) ($V_b$ = 50 mV, $I_t$ = 50 pA). Each bright elliptical feature represents one DCA molecule (see DCA chemical structure overlay). Vectors $\mathbf{a_1}$, $\mathbf{a_2}$ define a primitive unit cell of the crystalline self-assembly ($\|\mathbf{a_1}\|$ = 1.20 ± 0.02 nm; $\|\mathbf{a_2}\|$ = 0.99 ± 0.01 nm; $\sphericalangle(\mathbf{a_1}, \mathbf{a_2})$ = 53 ± 1°). (b) d$I$/d$V$ STS spectra taken across DCA domain boundary from bare Ag(111) (magenta curve) to DCA domain (cyan) in (a) (setpoint $V_b$ = -170 mV, $I_t$ = 50 pA). Spectra at the DCA centre within the molecular domain (cyan) show a step-like feature with an onset energy of ~120 mV, compared to the Ag(111) Shockley surface state onset (magenta) of ~-65 mV. Scattering of the latter by the molecular domain boundary results in oscillations in the LDOS (black arrows). d$I$/d$V$ STS spectrum taken at the DCA anthracene extremity [red cross in (a)] shows a peak-like feature at ~350 mV (red curve) which corresponds to the DCA LUMO.

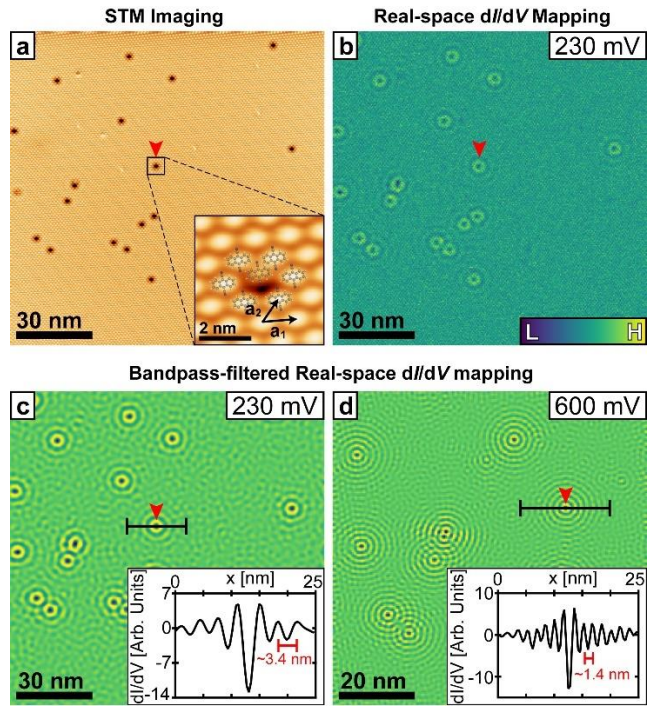

**Figure 3.** (a) Constant-current STM imaging of large self-assembled DCA molecular domain (124 × 124 nm$^2$) with tip-induced defects (setpoint $V_b$ = 230 mV, $I_t$ = 2 nA). Defects (dark depressions) are DCA molecules displaced from their initial position by interacting with the STM tip (see Methods). Inset: detail of defect surrounded by DCA molecules, with DCA chemical structure overlaid. (b) Constant-current d$I$/d$V$ STS mapping of region in (a); $V_b$ = 230 mV. (c)-(d) Bandpass-filtered constant-current d$I$/d$V$ STS maps of DCA molecular film with defects; $V_b$ = 230 and 600 mV, respectively. These maps include Fourier components for scattering wavevectors q with 1.4 < |**q**| < 2.3 nm$^{-1}$, with components due to the structural periodicity of the molecular film removed. Insets: d$I$/d$V$ line profiles across a single defect (black line) showing Friedel oscillations due to scattering of interface face, with wavelengths of ~3.4 and ~1.4 nm, respectively. Red tick marks the same defect within the molecular film.

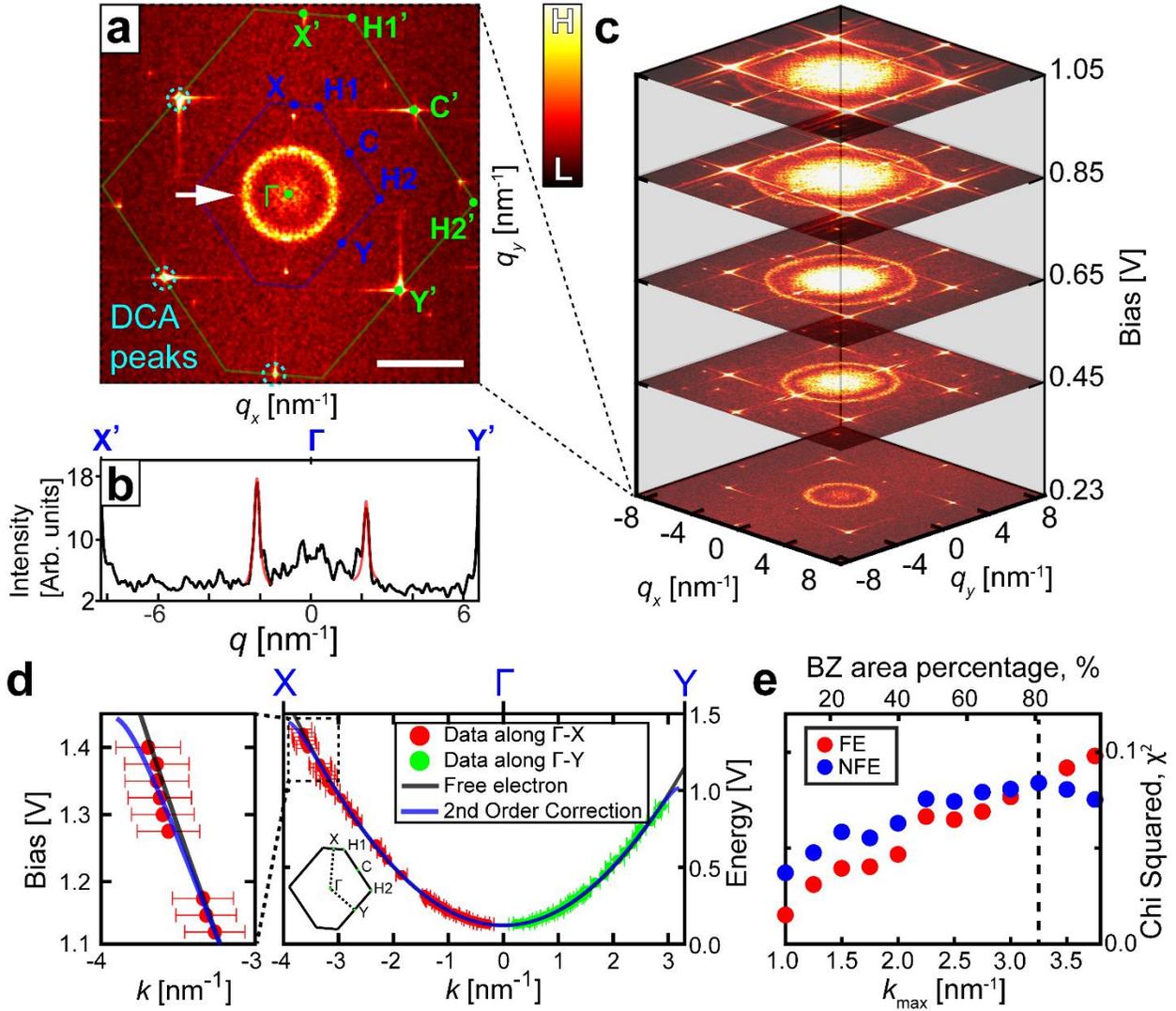

**Figure 4.** (a) Fourier-transform (FT) of d$I$/d$V$ map at $V_b$ = 230 mV in Fig. 3b, showing peaks given by the DCA film periodicity (cyan dashed circles). The radius of the ring-like feature (white arrow) centred at **q** = 0 nm$^{-1}$ is associated with the real-space LDOS modulation in Figs. 3c, d. Green (blue) lines indicate **q**-space (**k**-space; **q** = 2**k**) Brillouin zone (BZ) boundary of DCA molecular film, with high-symmetry points Γ, X', H1, C', H2', Y' (X, H1, C, H2, Y, respectively). Scale bar: 4 nm$^{-1}$. (b) FT intensity line profile along X′-Γ-Y′ in (a). The peak positions near |**q**| = 2 nm$^{-1}$ were extracted from Lorentzian fits (red curves), allowing us to determine the **q**- and **k**-space radius of the ring-like feature in (a) related to the IS scattering. (c) FTs of d$I$/d$V$ maps of defect-filled DCA

film at different bias voltages. (d) Energy dispersion, $E(\mathbf{k})$, of IS as a function of wavevector $k$, along X-Γ-Y, determined via Lorentzian fits of FT intensity profiles of ring-like scattering feature at different bias voltages in (c). FT intensity line profiles were averaged within an angle window (increasing gradually from ±1° at 120 mV to ±5° at 1.5 V) to increase signal-to-noise ratio. See Methods for details on error bars. Black curve: quadratic fit assuming free electron dispersion [Eq. (2)]. Blue curve: fit with $2^{nd}$ order corrections to the free electron dispersion [Eq. (1)]. Left: $E(\mathbf{k})$ detail for $-4 < k < -3$ nm$^{-1}$ along Γ-X. (e) $\chi^2$ for free-electron (red) and nearly-free-electron (blue) model fits of $E(\mathbf{k})$, as a function of $k_{max}$.

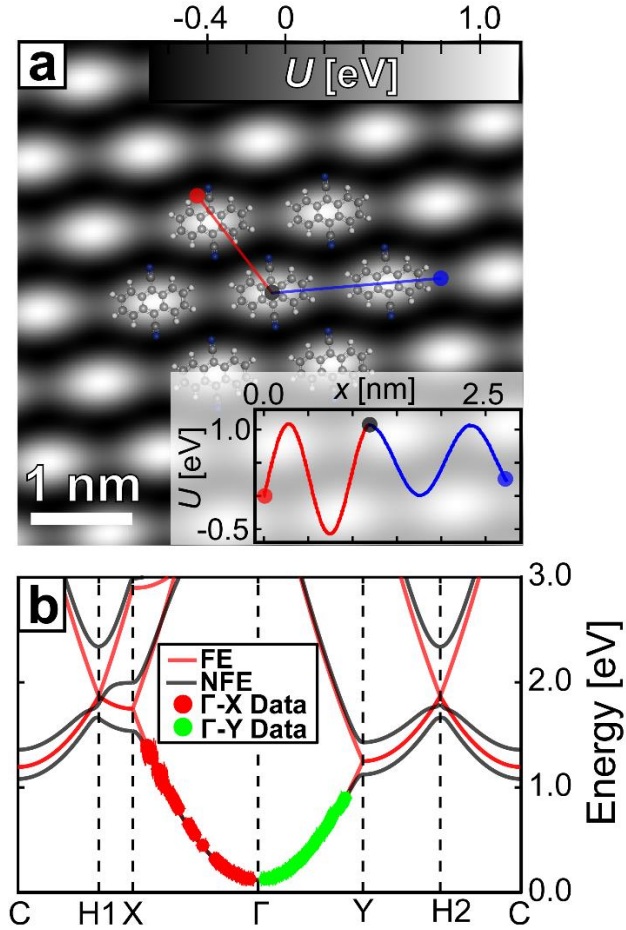

**Figure 5.** (a) Reconstructed upper bound of real-space potential, $U(\mathbf{r})$, resulting from the 2D DCA molecular film on Ag(111), for $U_{\pm \mathbf{G_1}} = 0.29$ meV, $U_{\pm \mathbf{G_2}} = 0.18$ meV and $U_{\pm (\mathbf{G_1}+\mathbf{G_2})} = 0.17$ meV, showing $\Delta U(\mathbf{r}) \approx 1.5$ eV. Origin $\mathbf{r} = \mathbf{0}$ corresponds to centre of DCA molecule (black circle). Inset: $U(x)$ along red and blue lines. (b) Calculated band structure $E(\mathbf{k})$ for the periodic molecular potential $U(\mathbf{r})$ in (a) using the nearly-free electron approximation (black) with an electron effective mass, $m^*/m_e = 0.385$. Calculated free electron band structure (i.e., for $U(\mathbf{r}) = 0$; red) shown for reference ($m^*/m_e = 0.385$). Red and green circles: $E(\mathbf{k})$ data along Γ-X Γ-Y, from Figure 4d.

AUTHOR INFORMATION

**Corresponding Author**

* agustin.schiffrin@monash.edu

ASSOCIATED CONTENT

**Supporting Information**. The Supporting Information is available free of charge on the XX website at DOI: XX. $E(\mathbf{k})$ along k-space high-symmetry directions of 2D DCA molecular film on Ag(111); FT-STS map at $V_b$ = 1.4V; Band structure calculation.

ACKNOWLEDGEMENTS

D.K. and J.H. acknowledge funding support from the Australian Research Council (ARC) Centre of Excellence in Future Low-Energy Electronics Technologies. A.S. acknowledge funding support from the ARC Future Fellowship scheme (FT150100426).

TOC GRAPHIC

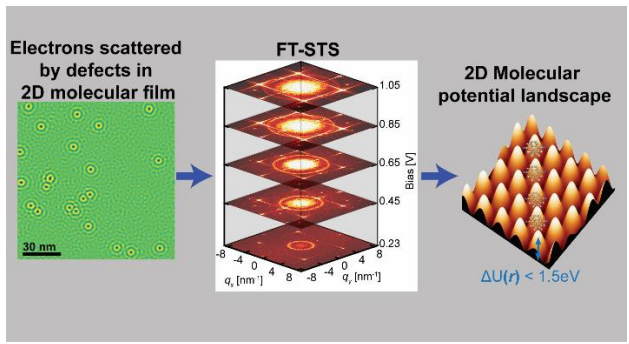

# Upper Bound Estimate of the Electronic Scattering Potential of a Weakly-interacting Molecular Film on a Metal

*Dhaneesh Kumar[†, §], Matthew Hendy[†], Jack Hellerstedt[†, §], Joanna Hewes[†], Shon Kolomoisky[†], Cornelius Krull[†, §], and Agustin Schiffrin[†, §, ]*](*)

[†]School of Physics & Astronomy, Monash University, Clayton, Victoria 3800, Australia

[§]ARC Centre of Excellence in Future Low-Energy Electronics Technologies, Monash University, Clayton, Victoria 3800, Australia

**Corresponding Author**

[*] agustin.schiffrin@monash.edu

## S1. $E(k)$ along k-space high-symmetry directions of 2D DCA molecular film on Ag(111)

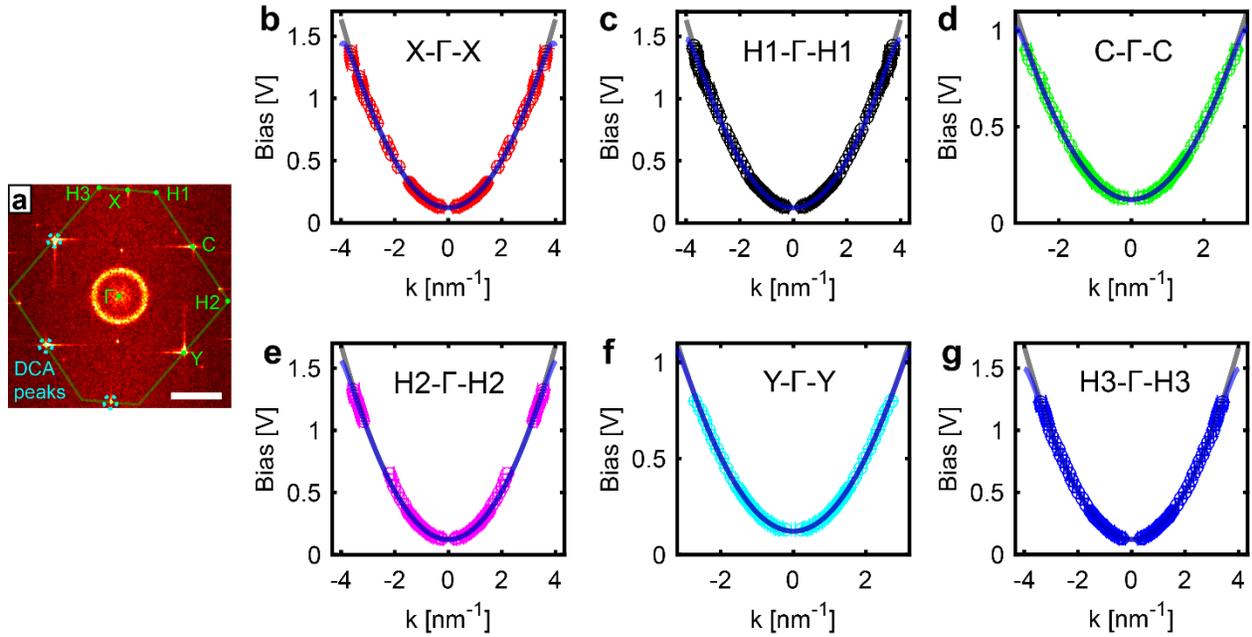

**Figure S1** (a) Fourier-transformed d$I$/d$V$ map at $V_b$ = 230 mV, reproduced from the main text Fig. 4a. Scale bar: 4 nm$^{-1}$. (b)–(g) Energy dispersion relations $E(\mathbf{k})$ of DCA/Ag(111) interface state along the indicated high-symmetry directions, according to high-symmetry points indicated in (a). Solid black curves: free electron dispersion [parabolic; Eq. (2) in main text] fit. Solid blue curves: nearly-free-electron dispersion fit with 2$^{nd}$ order corrections [Eq. (3) of main text].

## S2. FT-STS map at $V_b$ = 1.4V

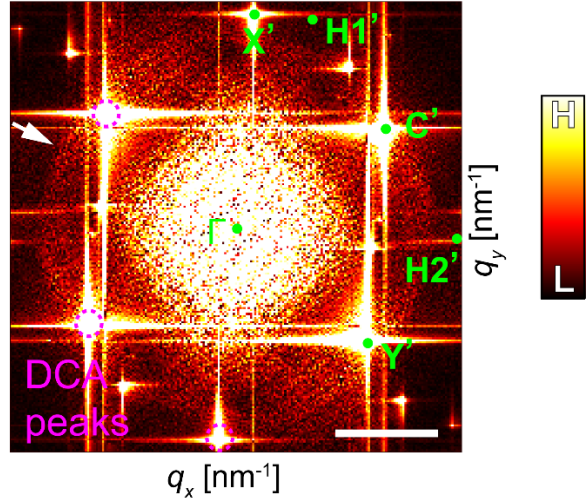

**Figure S2** Fourier-transform (FT) of d$I$/d$V$ map (for same DCA domain as in Fig. 3a of the main text) at $V_b$ = 1.4 V, showing a faint circular scattering feature (white arrow) centred at **q** = 0 nm$^{-1}$. Peaks corresponding to the DCA film periodicity (magenta dashed circles) and the DCA molecular film Brillouin zone (in **q**-space) high symmetry points Γ, X', H1', C', H2', Y' (green) are indicated. Scale bar: 4 nm$^{-1}$.

As noted in the main text, retrieval of the energy dispersion $E(\mathbf{k})$ for bias voltages $V_b \geq 1.4$ V is challenging due to: (i) severe attenuation of the d$I$/d$V$ scattering intensity at energies far above the IS onset ; (ii) overlap in the FT-STS maps of the structural Fourier peaks given by the 2D molecular film periodicity with the scattering intensity (given that $\mathbf{q} = 2\mathbf{k}$; see Figure 4a of the main text). This overlap challenges an accurate evaluation of $E(\mathbf{k})$ at energies where **k** is close to or at the BZ boundary. Note however that, in contrast with FT-STS d$I$/d$V$ maps at lower energies, the intensity of the circular scattering feature for $V_b$ = 1.4 V is not isotropic, and seems to decrease along Γ – X', Γ – H1' and in particular, Γ – H2'; this decrease in the FT-STS d$I$/d$V$ signal could be an indication of a gap opening.

## S3. Band structure calculation

We calculated the 2$^{nd}$-order-corrected, nearly-free-electron band structure, $E(\mathbf{k})$, for the interface state (Fig. 5 in main text) by solving the central equation given by Eq. (6) in the main text. In this calculation, we took into consideration the actual lattice geometry of the DCA film[1] (i.e., oblique 2D Bravais lattice with primitive unit cell vectors $\mathbf{a}_1$ and $\mathbf{a}_2$; see main text Fig. 3a), reflected in the reciprocal lattice vectors **G** by the relation:

$$\begin{pmatrix} G_{1x} & G_{2x} \\ G_{1y} & G_{2y} \end{pmatrix} = \frac{2\pi}{a_{1x}a_{2y} - a_{1y}a_{2x}} \begin{pmatrix} a_{2y} & -a_{1y} \\ -a_{2x} & a_{1x} \end{pmatrix} \qquad \text{(Eq. S3)}$$

Here, we outline this calculation. For simplicity, here we will solve main text Eq. (6) for an electron in a 1D periodic potential, $U(x)$, with lattice constant $a$, as an example. That is, the nearly-free-electron band structure, $E(\mathbf{k})$, in Fig. 5b of the main text results from a similar calculation, but performed in 2D, for a 2D scattering potential $U(\mathbf{r})$ [Eq. (4) in main text], and for the 2D reciprocal lattice vectors $\mathbf{G}$ in Eq. S3. Since $U(x)$ is periodic, the only non-zero Fourier coefficients, $U_G$, are those associated with reciprocal lattice vectors $G = 2\pi j/a = j\,b$ with $j \in \mathbb{Z}$. By symmetry, $U_G = U_{-G}$. Furthermore, we will consider, for simplicity, only $U_G = U_{\pm b}$, since these terms typically provide the largest contribution to the real-space potential $U(x)$. In its matrix form, Eq. (6) in the main text then becomes:

$$\begin{pmatrix} \cdots & \cdots & \cdots & \cdots & \cdots & \cdots & \cdots \\ \cdots & \lambda_{k+2b} - \epsilon & -U_b & 0 & 0 & 0 & \cdots \\ \cdots & -U_b & \lambda_{k+b} - \epsilon & -U_b & 0 & 0 & \cdots \\ \cdots & 0 & -U_b & \lambda_k - \epsilon & -U_b & 0 & \cdots \\ \cdots & 0 & 0 & -U_b & \lambda_{k-b} - \epsilon & -U_b & \cdots \\ \cdots & 0 & 0 & 0 & -U_b & \lambda_{k-2b} - \epsilon & \cdots \\ \cdots & \cdots & \cdots & \cdots & \cdots & \cdots & \cdots \end{pmatrix} \begin{pmatrix} \cdots \\ C_{k+2b} \\ C_{k+b} \\ C_k \\ C_{k-b} \\ C_{k-2b} \\ \cdots \end{pmatrix} = 0 \qquad \text{(Eq. S4)}$$

where $\lambda_k = \hbar^2 k^2 / 2m^*$, $\epsilon = E(k)$, and $C_k$'s are Fourier coefficients of the electron wavefunction. We only consider $|G| < G_{\max}$, with $G_{\max} = 2 \times 2\pi/a = 2b$:

$$\begin{pmatrix} \lambda_{k+2b} - \epsilon & -U_b & 0 & 0 & 0 \\ -U_b & \lambda_{k+b} - \epsilon & -U_b & 0 & 0 \\ 0 & -U_b & \lambda_k - \epsilon & -U_b & 0 \\ 0 & 0 & -U_b & \lambda_{k-b} - \epsilon & -U_b \\ 0 & 0 & 0 & -U_b & \lambda_{k-2b} - \epsilon \end{pmatrix} \begin{pmatrix} C_{k+2b} \\ C_{k+b} \\ C_k \\ C_{k-b} \\ C_{k-2b} \end{pmatrix} = 0 \qquad \text{(Eq. S5)}$$

which can be solved for the eigenenergies, $\epsilon$, by considering the determinant of the matrix:

$$\begin{vmatrix} \lambda_{k+2b} - \epsilon & -U_b & 0 & 0 & 0 \\ -U_b & \lambda_{k+b} - \epsilon & -U_b & 0 & 0 \\ 0 & -U_b & \lambda_k - \epsilon & -U_b & 0 \\ 0 & 0 & -U_b & \lambda_{k-b} - \epsilon & -U_b \\ 0 & 0 & 0 & -U_b & \lambda_{k-2b} - \epsilon \end{vmatrix} = 0 \qquad \text{(Eq. S6)}$$

Solving Eq. (S6) for $k$'s all over the Brillouin zone allows, then, for the construction of the band structure $E(k)$. In the main text, we limited ourselves to the first five energy bands of the 2D band structure, given the truncation of the matrix form of Eq. (6) of the main text. By considering a larger $G_{\max}$, we can account for higher energy bands. This procedure can be generalised to any periodic potential of any dimensionality.